# A New Approach to Overcoming Zero Trade in Gravity Models to Avoid Indefinite Values in Linear Logarithmic Equations and Parameter Verification Using Machine Learning

Mikrajuddin Abdullah

*Department of Physics, Bandung Institute of Technology, Jalan Ganesa 10, Bandung, 40132, Indonesia*

E-mail: mikrajuddin@gmail.com

Abstract

The presence of a high number of zero flow trades continues to provide a challenge in identifying gravity parameters to explain international trade using the gravity model. Linear regression with a logarithmic linear equation encounters an indefinite value on the logarithmic trade. Although several approches to solving this problem have been proposed, the majority of them are no longer based on linear regression, making the process of finding solutions more complex. In this work, we suggest a two-step technique for determining the gravity parameters: first, perform linear regression locally to establish a dummy value to substitute trade flow zero, and then estimating the gravity parameters. Iterative techniques are used to determine the optimum parameters. Machine learning is used to test the estimated parameters by analyzing their position in the cluster. We calculated international trade figures for 2004, 2009, 2014, and 2019. We just examine the classic gravity equation and discover that the powers of GDP and distance are in the same cluster and are both worth roughly one. The strategy presented here can be used to solve other problems involving log-linear regression.

# 1 Introduction

The gravity model is widely used for analyzing a wide range of international trade because of its apparent good performance in representing trade flows (Martin and Pham, 2020). Because of the advancement of its microeconomic foundations, the gravity equation model has recently gained increased attention (Xiong Chen, 2014). In its traditional form, the gravitational equation for international trade is given by (Anderson, 2011; . Chaney, 2018)

$$F_{ij} = K G_i^\delta G_j^\delta R_{ij}^\beta \qquad (1)$$

where $F_{ij}$ represents the volume of trade between countries i and j, $G_i(G_j)$ denotes the GDP of country i(j), and $R_{ij}$ denotes the distance between countries i and j (generally, the distance between capital cities) (Wall, 1999). We use the same exponents for $G_i$ and $G_j$ as the other authors (Gul and Yasin, 2011; Sohn , 2001; Shahriar *et al*., 2019; Battersby and Ewing, 2005). Equation (1) is generally converted into logarithmic form as follows

$$\ln F_{ij} = \ln K + \delta \ln(G_i G_j) + \beta \ln R_{ij} \qquad (2)$$

As previously reported, if $F_{ij} = 0$ (zero flow) problems arise in Eq. (2) because it results in $\ln F_{ij} \to \infty$ making linear regression impossible. Many methods for overcoming the zero low have been proposed. One method is to exclude trades with $F_{ij} = 0$ from consideration. By removing zero flow trades, the amount of data used for parameter estimation is reduced. For information, the number of country pairs with no trades is nearly equal to the number of country pairs with trades. This is due to the fact that each country only trades with limited number of countries. Only large GDP countries trade with more than 200 other countries.

Another method proposed by the researchers to include countries with zero flow trades in the calculation is to add one to the trade to become $F_{ij} + 1$ or use a tobit approximation by adding a small constant $a$ to all trade pairs of countries to become $F_{ij} + a$ (Eaton and Tamura, 1994; Bellego, Benatia, and Pape, 2022). Trades with large values do not change when one or $a$ is added ($F_{ij} + 1 \approx F_{ij}$ and $F_{ij} + a \approx F_{ij}$), but trades with zero values do not produce infinite values when converted to logarithms. However, this procedure has not been supported by strong reason.

Another approach is to use formulations of the zero-inflated negative binomial and zero-inflated Poisson models (Lambert, 1992; Burger, van Oort and Linders, 2009). In this model, different distribution functions explain null and nonzero data (Heckman, 1979). Di Vece et al. reported a detailed description of the two models (Di Vece, Garlaschelli and Squartini, 2022).

The next approach is to estimate parameters directly in the multiplication form equation rather than using Eq. (2) (Santos Silva and Tenreyro, 2006). Because no infinite value is generated, this approach can include all zero and non-zero trades. Santos Silva and Tenreyro (2006, 2022) demonstrated using Monte Carlo simulations that traditional estimators based on the linearized logarithm formalism can produce severely biased parameter estimates. According to several reports, the estimated value of the gravity parameters varies greatly. Santos Silva and Tenreyro used the Poisson Pseudo Maximum Likelihood (PPML) estimator as the preferred estimator for dealing with these problems (Santos Silva and Tenreyro, 2006, 2022). This step, however, is more difficult than linear regression, which is widely used for parameter estimation. Furthermore, Martin and Pham (2020) found that when zeros are not random outcomes, the PPML estimates are severely biased. Therefore, there is a lack of agreement on the optimum strategy to address those zeros in log-linear and log-log models, as seen by the numerous diverse ways utilized in recent major papers.

One of the most noticeable issues in using the gravity model to explain international trade is the wide range of parameters $\delta$ and $\beta$. These parameters were reported to have substantially diverse values by different researchers. Gul and Yasin (2011), for example, observed $\delta = 1.65$ for trade between Pakistan and NAFTA or Latin America, whereas Almog, Bird, and Garlaschelli (2019) reported $\delta$ values ranging from 0.67 to 0.91. For the $\beta$ parameter, Gul and Yasin (2011) reported $\beta = -7.99$ for trade between Pakistan and the Middle East, while Frenkel (1997) reported $\beta = -0.732$. Other researchers reported values in between the ranges. Although there is some fluctuation in the calculated parameters, it should not be as extensive as the present researchers found. Therefore, the following question is, within what range should the values of the parameters $\beta$ and $\delta$ be?

The approach that we will describe in this paper is divided into two stages. The first stage is to estimate the parameters in the traditional gravity equation on a global scale. The second stage is to make local adjustments at the end to bring the estimates closer to the actual data. When we use all trade data from all countries, the parameters we get are global parameters. Some global parameters are also local parameters, and others must change when the equation is applied locally. We assume the gravity model is based on the fact that the volume of trade between two countries is affected by both GDP and distance on the same power. This is consistent with Chaney's finding that the values of $\beta \approx -1$ and $\delta \approx 1$ have remained stable over a century (Chaney, 2018) so it is very appropriate if we choose them as a global parameter. Other factors, however, are country-specific and differ between pairs of countries. This factor is represented by the constant $K$ in Eq. (1). We treat the $K$ the same way for all countries in global estimation until we get the best estimate for $\delta$ and $\beta$. Following that, we adjust the $K$ to be unique to each country pair. This approximation can be used to represent a modified gravitational equation of the general form (Nijkamp and Ratajczak, 2021)

$$F_{ij} = K G_i^\delta G_j^\delta R_{ij}^\beta X_{ij} Y_{ij} Z_{ij} \ldots \tag{3}$$

where $X_{ij}$, $Y_{ij}$, and $Z_{ij}$ are other influential factors such as population, per capita income, and border factor. This equation, like the traditional gravity equation, can be written as $F_{ij} = K_{ij} G_i^\delta G_j^\delta R_{ij}^\beta$, except that $K_{ij}$ is no longer a constant.

In this paper, we first consider $K_{ij}$ as a global constant to estimate global parameters $\delta$ and $\beta$, in the form of $F_{ij} = \overline{K_{ij}} G_i^\delta G_j^\delta R_{ij}^\beta$. Then, using Eq. (2), we apply the linear regression. Following the determination of the global parameters $\overline{K_{ij}}$, $\delta$, and $\beta$, we determine the specific $K_{ij}$ for each country pair in order to obtain a specific equation for each country pair.

## 2 Method

We propose a new method for estimating parameters that involves all trades while still using linear regression on logarithmic equations. The steps are as follows:

### 2.1 Parameters Estimation

a) Filling in dummy values for zero flow trades

We prevent the appearance of the indefinite value by replacing the zero flow trade with a dummy value. The nearest neighbor linear regression method, as described below, is used for this purpose.

According to Eq. (1), $F_{ij}$ is a linear function of $G_i^\delta G_j^\delta R_{ij}^\beta$. If $F_{ij} = 0$, we replace it with a dummy value resulting from linear regression using $F_{i'j'}$ with a value of $G_{i'}^\delta G_{j'}^\delta R_{i'j'}^\beta$, adjacent to $G_i^\delta G_j^\delta R_{ij}^\beta$. We select the $M$ values $G_{i'}^\delta G_{j'}^\delta R_{i'j'}^\beta$ that is closest to the $G_i^\delta G_j^\delta R_{ij}^\beta$. $F_{ij}$ is estimated by plugging $G_i^\delta G_j^\delta R_{ij}^\beta$ into the regression equation. The number of neighbors with the closest values involved can be arbitrary, but not excessively so that the values of $G_{i'}^\delta G_{j'}^\delta R_{i'j'}^\beta$ used are not too far away from $G_i^\delta G_j^\delta R_{ij}^\beta$. This procedure is carried out for all $F_{ij}$ values that are zero.

Figure 1 shows how to find a dummy value to replace $F_{ij}$ which is zero after sorting the data, by increasing $G_i^\delta G_j^\delta R_{ij}^\beta$. We demonstrate linear regression using the four neighbors with the closest $G_i^\delta G_j^\delta R_{ij}^\beta$ values, but we can use more nearest neighbors in the simulation. Data number two, for example, has $F_{ji} = 0$. We replace this data with a linear regression involving data points 1, 3, 4, and 5. The regression outcome is a dummy $F_{ij}$ value associated with data 2'. This new $F_{ij}$ alue will be used in the logarithmic linear regression equation, which includes all data, including the original data ($T_{ij} \neq 0$) and the dummy $F_{ij}$. Similarly, data 7 is replaced with the regression results obtained from data 6, 8, 9, and 11. $F_{ij}$ is the result, and it is related to data number 7'. And so on until all zero trading data becomes non-zero..

We enter arbitrary initial values for the parameters $\delta = \delta_1$ and $\beta = \beta_1$ to calculate $G_i^\delta G_j^\delta R_{ij}^\beta$ for all country pairs. The process is repeated on all zero $F_{ij}$ resulting in a $F$ matrix with no zero elements other than the diagonal elements $F_{ii}$. As a result, the values of $\ln F_{ij}$ (for $i \neq j$) become finite.

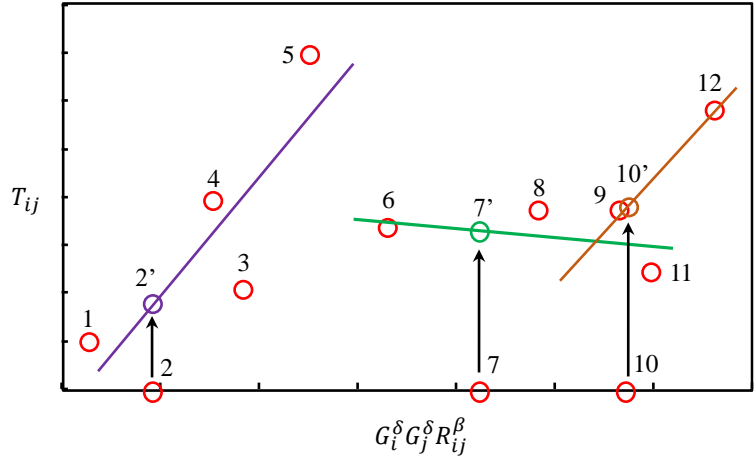

**Fig. 1** Illustration of a linear regression process to obtain a dummy value as a substitute for a zero $T_{ij}$

b) Variable Normalization

The trade value of $T_{ij}$ ranges from millions to trillions of dollars, $GG$ can be worth hundreds of trillions of dollars, and $R$ can be worth tens of thousands of kilometers. We normalize these quantities to the same order for use in regression. We normalize the values to the total, i.e.

$$t_{ij} = \frac{F_{ij}}{\sum F_{ij}} \tag{4}$$

$$gg_{ij} = \frac{G_i G_j}{\sum G_i G_j} \tag{5}$$

$$r_{ij} = \frac{R_{ij}}{\sum R_{ij}} \tag{6}$$

c) Converting Matrices to Vectors

$F_{ij}$, $G_i^\delta G_j^\delta$, and $R_{ij}$ are $N \times N$ matrices, where $N$ is the number of countries involved in the calculation. These variables are symmetrical variables, such as,

$$F_{ij} = Ex_{ij} + Im_{ij} \tag{7}$$

where $Ex_{ij}$ is the amount of exports from country i to country j and $Im_{ij}$ denotes the value of country i's imports from country j. Because $Ex_{ij} = Im_{ji}$, we get $F_{ij} = F_{ji}$. This variable is similar to the undirected (symmetrized) version of the weighted trade matrix, $w_{ij} = (Ex_{ij} + Ex_{ji})/2$, used by Di Vece *et al.* (2022). Because $Ex_{ji} = Im_{ij}$, we get $w_{ij} = (Ex_{ij} + Im_{ij})/2 = F_{ij}/2$. With this property, we only need to use half diagonal data from the matrices above.

For $F_{ij}$, many authors have used export-only or import-only values (Nijkamp and Ratajczak, 2021; Paas, 2000). As a result, $F_{ij}$ is not a symmetrical quantity, which means that the "attraction force" of country i to country j is not the same as the "attraction force" of country j to country i. This property contradicts the law of gravity, which states that the gravitational force exerted by country i on country j equals the attractive force exerted by country j on country i. When exports and imports are separated, there is sometimes a significant difference in elasticity between the two equations. For example, Paas reported $\beta = -1.931$ in the export equation and $\beta = -0.851$ in the import equation (Paas, 2000). This is one of the logical reasons why $F_{ij}$ is associated with total export and import. We are only concerned with the top (or bottom) off-diagonal elements of matrix $F$ with this property. The top off-diagonal element is used in the next calculation.

Following that, the top off-diagonal element of $F$ is vectorized. $N(N-1)/2$ is the number of elements in the vector. Combining Eqs. (2), (4)-(6) yields the following equation

$$\ln \tilde{t} = \ln K + \delta \ln \widetilde{gg} + \beta \ln \tilde{r} \qquad (8)$$

where the wavy cap sign denotes a vector of length $N(N-1)/2$.

### d) Linear Regression

Following that, we perform linear regression on Eq. (8) to obtain new parameters $\delta_2$ and $\beta_2$. Then, using these new parameters, we repeat steps (a)-(d). This stage will change the dummy value as a substitute for $F_{ij}$, which has a value of zero, whereas non-zero $F_{ij}$ has no effect. The process is repeated until the conditions $|\delta_2 - \delta_1| \leq \epsilon_1$ and $|\beta_2 - \beta_1| \leq \epsilon_2$ are met, where $\epsilon_1$ and $\epsilon_2$ are very small numbers. A flowchart for determining the parameters $\delta$ and $\beta$ is shown in Fig. 2.

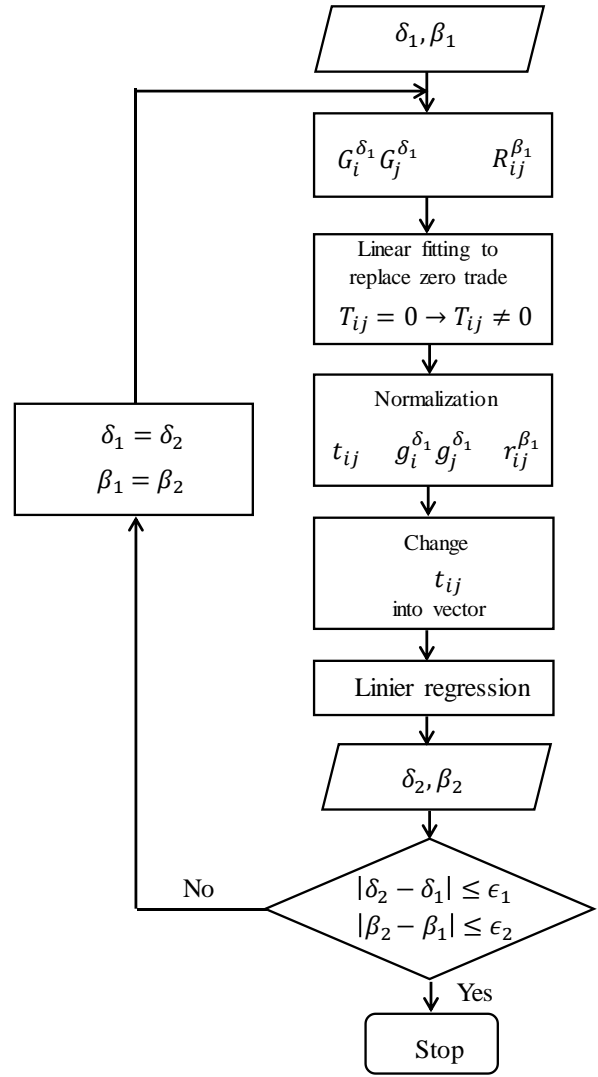

**Fig. 2** A flowchart for determining the parameters $\delta$ and $\beta$.

e) Post Adjustment

The final results yield three parameters: intercept $(\ln K)$, power of GDP $(\delta)$, and power of distance $(\beta)$. The final equation should satisfy Eq. (1). However, when the obtained equations are compared to real data, a large deviation is usually found. If $K$ satisfies the equation $K = F_{ij}/G_i^\delta G_j^\delta R_{ij}^\beta$, the regression equation is the same as the real, but it is not a constant.

If we replace $K = K'/\kappa_{ij}$ with $K'$ being constant and $\kappa_{ij}$ is not constant, we get the equation

$$F_{ij} = \frac{K\prime}{\kappa_{ij}} G_i^\delta G_j^\delta R_{ij}^\beta \tag{9}$$

Equation (9) is analogous to the coulomb equation of force between the electric charge in the dielectric material and the "dielectric constant" $\kappa_{ij}$,

$$F_{ij} = \frac{(1/4\pi\epsilon_0)}{\kappa_{ij}} Q_1 Q_2 R^{-2} \tag{10}$$

where $Q_i$ is the electric charge and $\epsilon_0$ is the vacuum permittivity. We contend that international trade is more akin to the coulomb equation (Eq. (10)) than the gravitational equation (Eq. (1)).

The coulomb equation and gravity behave similarly in that force is proportional to the mass/charge product and inversely proportional to the square of the distance. The main distinction between the two equations is that the gravitational equation always produces the same force as long as the product of mass and distance is the same. Furthermore, in the gravitational equation, the force cannot be zero if the distance is not too large. In contrast, even though the multiplication of charges and distances is the same in the coulomb equation, the resulting force can be different. If the dielectric constant varies, the resulting force can vary. In fact, if the dielectric constant approaches infinity, the resulting force can approach zero over relatively short distances. Observational data on trade between countries are closer to predictions of the coulomb force than predictions of the gravitational force because there is no trade even when the distance between two countries is not too large. Other than GDP and distance, variables that affect trade can be aggregated into a single "dielectric constant." We conclude that the coulomb force model simplifies the international trade equation.

Furthermore, we can write Eq. (10) in the form

$$F_{ij} = K\prime G_i^\delta G_j^\delta T_{ij}^\beta \tag{11}$$

where

$$T_{ij} = \frac{R_{ij}}{\kappa_{ij}^{1/\beta}} \tag{12}$$

which is the "bilateral distance" between countries i and j, a dyadic variable. A country pair with the same spatial distance can have a different "bilateral distance" if the "dielectric constant" is different. The dielectric constant includes factors such as diplomatic relations between countries, war conditions, and so on. The "bilateral distance" can change over time, so it is a function of time in general, whereas spatial distance is time independent. As a result, $dR_{ij}/dt = 0$, whereas, in general, $dT_{ij}/dt \neq 0$.

For example, the European Union's (EU) trade with Russia has decreased significantly as a result of Russia's invasion of Ukraine since February 2022 (Eurostat, 2023). Between February and September 2022, Russia's share of extra-EU imports fell from 9.5% to 5.3%. During the same time period, extra-EU exports to Russia fell from 4.0% to 1.8%. During the same time period, Russia's GDP increased from 1.779 trillion USD (World Bank, 2022) in 2021 to 2.133 trillion USD (Wikipedia, 2023). in 2022, while the EU's GDP falls slightly from 17.177 trillion USD (World Bank, 2022) in 2021 to 16.6 trillion USD (Wikipedia, 2023). However, the Russian and EU GDPs multiplication increased from 30.56 in 2021 to 35.41 (trillion USD)$^2$ in 2022. This phenomenon contradicts the gravity theory of international trade, which states that trade volume is directly proportional to a certain power of GDP multiplication and inversely proportional to a certain power of distance. Because the distance between Russia and the EU remains constant, if we apply the theory of gravity, trade volumes between Russia and the EU should increase in 2022.

The distance between capital cities i and j is calculated by the equation (Abdullah, 2016)

$$R_{ij} = 2R \arcsin\left(\sqrt{\sin^2\left(\frac{\varphi_i - \varphi_i}{2}\right) + \cos\varphi_1 \cos\varphi_2 \sin^2\left(\frac{\lambda_i - \lambda_i}{2}\right)}\right) \quad (13)$$

where $R = 6{,}371$ km is the radius of the earth, $\varphi_i$ represents the latitude capital city i, $\varphi_j$ represents the latitude capital city j, $\lambda_i$ represents the longitude capital city i, and $\lambda_j$ represents the longitude capital city j.

## 2.2 Data Sources

International trade data was retrieved from World Integrated Trade Solution (WITS) (2023), data on GDP was obtained from the World Bank (2022), and distances between countries' capitals were obtained from Nomad (2023). We can also record each one individually by typing "latitude longitude city's name" into Google search. We estimate data for the years 2004, 2009, 2014, and 2019, i.e. data prior to the Covid-19 pandemic. The countries included in the calculations were chosen as follows. For a specific year, such as 2019, we select countries that have WITS trade data for 2019. This information is not available in all countries. It is possible that a country will have data in one year but not in another. This implies that the countries involved in 2019 and 2014 may differ slightly. After determining which countries have trade data, we record the country's GDP using World Bank data. The World Bank provides very complete GDP data, so there are more country data on the World Bank than data on WITS-traded countries. We only select countries from the World Bank data that are listed in the WITS data. If a country is recorded in WITS but its GDP is not recorded in World Bank data, we seek the country's GDP data from other sources. There are only one or two countries missing from the World Bank data in this case. We do the same for the capital city latitude and longitude coordinate data, where we only have the capital cities of the countries in the WITS data. As a result, the number of countries involved in the calculation is approximately 150-160 (based on WITS data).

## 3 Results and Discussion

### 3.1 Estimation of Parameters

We begin by examining the iteration process, particularly the convergence process of the parameters $\delta$ and $\beta$. We enter various initial values for $\delta$ and $\beta$, namely $(\delta_1 = 1, \beta_1 = -1)$, $(\delta_1 = 1, \beta_1 = -0.5)$, and $(\delta_1 = 2, \beta_1 = -1)$. Figure 3 depicts the evolution of these parameters' values as the number of iterations increases. In iteration, we calculate the difference between two consecutive $\delta$ and two consecutive $\beta$ and the iteration is considered complete if the difference is less than 0.001. Iterations were performed more than 20 times in 2004 to converge the values of $\delta$ and $\beta$ when the initial values $\delta_1 = 1$, and $\beta_1 = -1$ were used. Iteration occurs faster (less than 20 times) for pairs of different and values $\delta$ and $\beta$, and in some cases only three times. We can also see that the convergence values for the pairs are always the same for the same year, regardless of the initial values assigned to the pairs $\delta$ and $\beta$. However, as shown in Table 1, the values in different years differ slightly. The value of the $\delta$ parameter ranges from 0.983 (2014) to 1.218 (2019) while the value of the $\beta$ parameter ranges from -1.374 (2019) to -1.079 (2009). Taking the average, we get $\bar{\delta} = 1.053$ and $\bar{\beta} = -1.186$. This value is stable and very close to that stated by Chaney (2018). According to Chaney, Head and Mayer (2014) examined the distance elasticity of hundreds of published papers and found that the mean distance elasticity for structural estimates was $\beta \approx -1.1$ (median -1.14) and for all estimates was $\beta \approx -0.93$ (median -0.8).

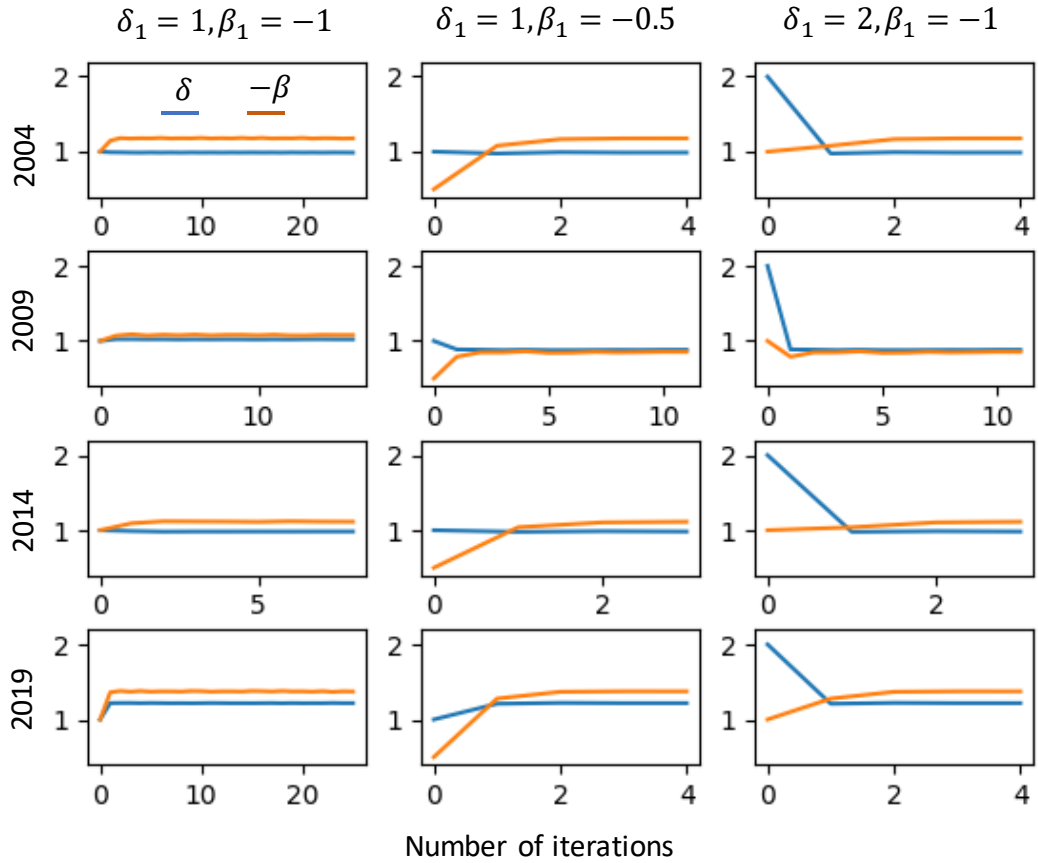

**Fig. 3** Evolution of $\delta$ and $\delta$ parameter pairs as a function of iteration count. The figure shows the different $\delta_1$ and $\beta_1$ pairs from left to right, while the top to bottom shows a different year

**Table 1** The estimation of the parameters $\delta$ and $\beta$ in 2004, 2009, 2014, and 2019

| Year | $\delta$ | $\beta$ |
| --- | --- | --- |
| 2004 | 0.988 | -1.178 |
| 2009 | 1.022 | -1.079 |
| 2014 | 0.983 | -1.113 |
| 2019 | 1.218 | -1.374 |

Several reports stated that the value of the $\delta$ are 0.728 (Sohn, 2001) and $0.67 - 0.91$ (Almog, Bird and Garlaschelli, 2019). Gul and Yasin (2011) have reported the following $\delta$ parameters for Pakistan's trade with other countries: Pakistan vs all countries (0.89), Pakistan vs EU (0.97), Pakistan vs ASEAN (0.65, ), Pakistan vs SAAR-ECO (0.61, ), Pakistan vs Middle East (0.92, ), Pakistan vs Far East (0.66), and Pakistan vs NAFTA and Latin America (1.65).

Reports on various $\beta$ paremeter values are $-0.72$ (Bergstrand, 1985), $-0.942$ (Sohn, 2001) and $-0.924$ (Wall, 1999). Gul and Yasin (2011) have reported the various values of β for Pakistan's trade with EU ($\beta = -1$), Pakistan's trade vs ASEAN ($\beta = -0.81$), Pakistan's trade vs SAAR-ECO ($\beta = -0.35$), Pakistan's trade vs Middle East ($\beta = -7.99$), Pakistan's trade vs Far East ($\beta = 0.77$), and Pakistan's trade vs NAFTA and Latin America ($\beta = -1.93$). Other reported parameters are β = -0.956 (Viorică, 2012), β = -1 (Chaney, 2013), β = -0.705 (Yuniarti, 2009), and β = -1.281 (Binh, Duong and Cuong, 2009), $\beta = -0.76$ (Linnemann, 1966), $\beta = -0.732$ (Frankel, 1997), $\beta = -0.942$ (Garman and Gilliard, 1999).

Figure 4 shows the original trade vector $\tilde{t}$ elements (a) and the elements after using linear regression to replace the zero elements (b). We use 2014 trade data and show only the first 500 elements of vector $\tilde{t}$. Figure 4(a) shows a lot of zero flow elements. There are no more zero-value elements in Fig. 4(b) because they have been replaced by the dummy regression results. Figures 4(a) and 4(b) also show that the non-zero elements in (a) do not change in (b).

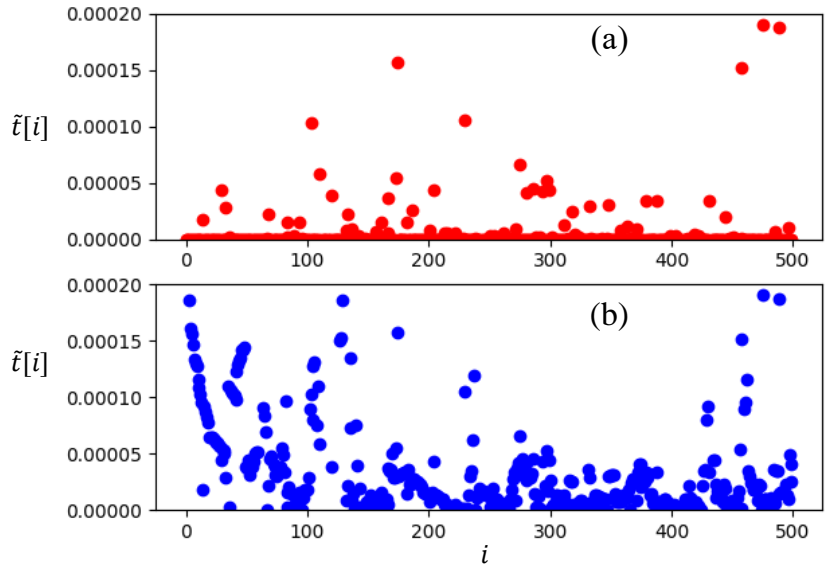

**Fig. 4** An example of the original $\tilde{t}$ trade vector element (a) and after the zero element was replaced with linear regression (b). We use 2014 trade data and show only the first 500 elements of vector $\tilde{t}$.

## 3.2 Clustering by Machine Learning

To validate the results of the simulations performed on the reported data, we divide the reported data into a number of clusters. We performed *K-means* clustering on the data reported by the researchers (Martin and Pham, 2020; Chaney, 2018; Wall, 1999; Gul and Yasin, 2011; Shahriar et al., 2019; Battersby and Ewing, 2005; Santos Silva and Tenreyro, 2006, 2022; Paas, 2000; Linders and de Groot, 2006; Matias, 2004; Haryadi and Hodijah, 2019; Kalirajan, 2007; Alleyne and Lorde, 2014; Effendi, 2014; Baldwin and Taglioni, 2011), which is one of the classification methods in machine learning. The elbow method is used to determine the number of clusters by looking for inertia that is nearly constant but has the fewest clusters. Figure 5(a) depicts the inertia as a function of the number of clusters. Based on the graph, we determine that the optimal number of clusters is four because it approaches a constant value when the number of clusters is increased.

We also compute a dendrogram for the same data, as shown in Fig. 5(b). The figure shows that dividing data into four clusters is the best option. Because there are four clusters, the K-*means* algorithm generates four dependent variables with indices [0,1,2,3].

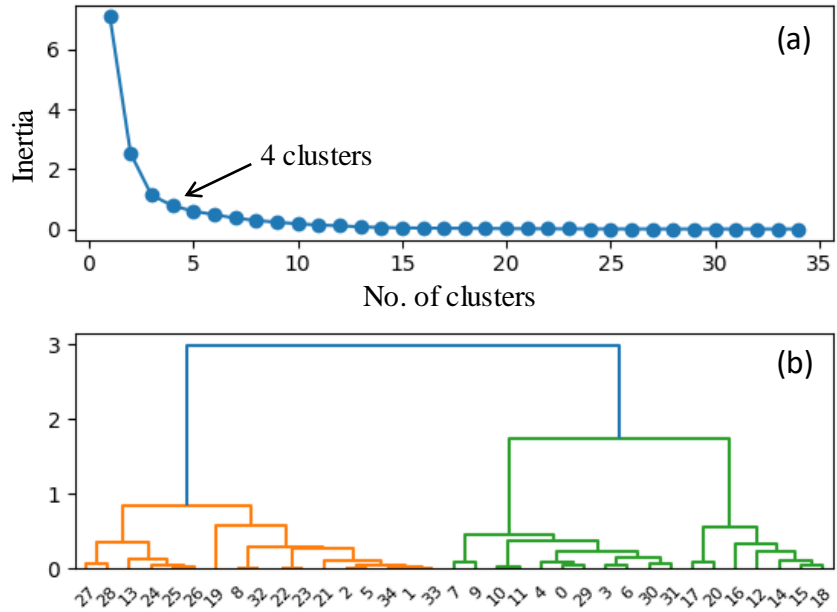

**Fig. 5** (a) Inertia vs. cluster number using the elbow method and (b) Dendrograms of the data.

The researchers' observational data and the dependent variable were divided into 80% training data and 20% test data. *The K-nearest neighbor* algorithm is trained on training data and tested on test data. As shown in Fig. 6(a), the test results are presented as a confusion matrix. We can easily see from the graph that the accuracy = 1.

Then we look at the cluster diagram to see where the simulated data is located. To determine the location of the simulated data, we use the results of the K-*nearest neighbor* algorithm. In Fig. 6(b), the four data sets obtained from simulation are denoted by a cross. All of the simulation data appears to be in the same cluster. The simulation results are also in the same cluster as Chaney's data where $\delta = 1$ and $\beta = -1$, which have been thought to be stable for nearly a century (Chaney, 2018). all calculation procedures employ the *sklearn library* in Python programming. We conclude from the results that the simulation produced good results because all of the data for all tangs is in one cluster, indicating that the parameters are stable.

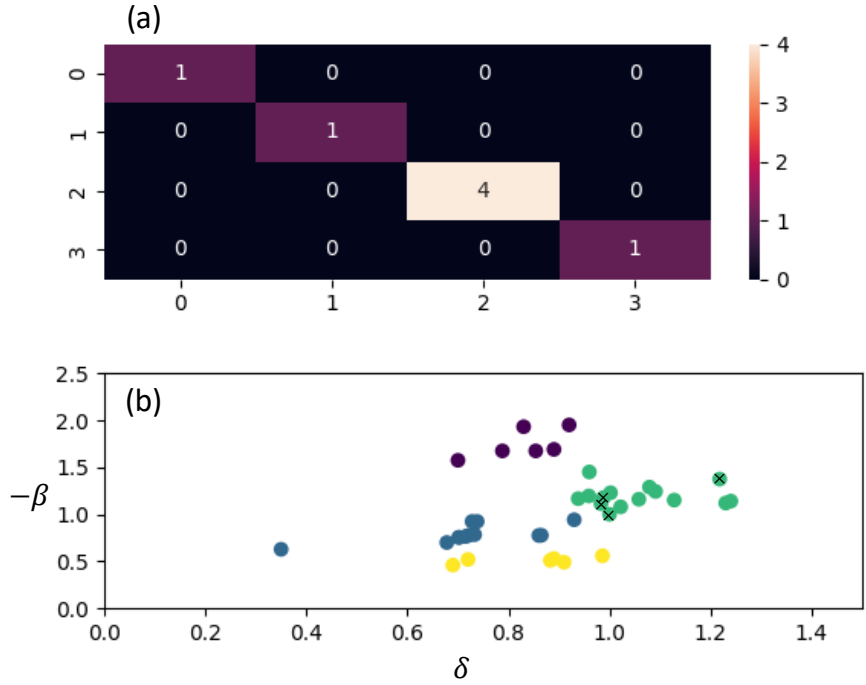

**Fig. 6** (a) The confusion matrix between the predicted and the test data (b) *K-means* clustering and *K-nearest neighbor* classification (using four nearest neighbors) (Martin and Pham, 2020; Chaney, 2018; Wall, 1999; Gul and Yasin, 2011; Shahriar et al., 2019; Battersby and Ewing, 2005; Santos Silva and Tenreyro, 2006; Paas, 2000; Linders and de Groot, 2006; Matias, 2004; Haryadi and Hodijah, 2019; Kalirajan, 2007; Alleyne and Lorde, 2014; Effendi, 2014; Baldwin and Taglioni, 2011)

**3.3 Final Notes**

The method proposed here is not only utilized to estimate the parameters of the gravity equation, but it may also be used to address analogous scenarios in other issues. We will have no difficulty completing the logarithmic procedure on the zero-valued dependent variable because the value will be replaced by the result of a linear fitting using the nearest pair of values. Furthermore, the number of iterations required to achieve convergence is not excessive, allowing the total process to run quickly. We expect the iteration process to be completed quickly in similar cases, such as for estimating plant biomass (Baskerville, 1972), rape conviction (Pugh, 1983), and others.

Falter and Perignon (2000) used the log attendant as the dependent variable to examine demand for French Premiere Division football matches played during the 1997/1998 season. If the precise number of attendees for a match is not recorded, we can use the method described here to approximate the number of attendees for that match. Hingley and Park (2015) used a lognormal model to forecat annual and quarterly total patent filings at the European Patent Office (EPO). In the event that some of the dependent variable data is lost, the approach provided here can be utilized to identify the dummy of the dependent variable by performing local linear regression.

An additional point to consider is trade flow, which measures the flow of goods expressed in currency. The gravitational equation, which is used to explain international trade, is a physics idea. The quantity that represents flow in physics is generated by the presence of potential. The magnitude of the flow created is proportional to the potential and is determined by the medium in which the flow occurs. Even if the potential is the same, the resulting flow may differ depending on the nature of the medium through which the flow is carried. Electric current exists due to the presence of an electric potential, and the magnitude of the flow is determined by the potential and resistivity of the material. The heat flow is determined by the heat potential (temperature difference), and the magnitude of the flow is determined by the potential and the material's thermal conductivity. Finally, fluid flow is affected by potential (pressure difference), and the magnitude of flow is affected by potential and material permeability. We can maintain a universal gravitational equation for all pairs of states if we consider that

$$P_{ij} = K G_i^\delta G_j^\delta R_{ij}^\beta \qquad (14)$$

as the **trade potential** of countries i and j, not as the trade flow of countries i and j. The **trade flow** between countries i and j is denoted as

$$F_{ij} = \frac{1}{\kappa_{ij}} P_{ij} \qquad (15)$$

Equation (15) is identical to the electric current equation, $I = (1/R)V$, Darcy's law, $q = -(k/\mu L)\Delta p$, and heat flux equation, $q = -k\nabla T$. Equation (15) is identical to the coulomb force equation.

## 4 Conclusion

We demonstrated a two-step method for predicting gravity parameters for international trade using basic linear regression. The method we present addresses the zero flow problem, which makes logarithmic linear regression problematic. Estimated parameters remained consistent from 2004 to 2019 and are close to values accepted by various studies. The clustering test from machine learning using K-*means* and K-*nearest neighbor* reveals that all parameters are in the same cluster As a result, the allowable values for the parameters $\delta$ and $\beta$ are all within one. This strategy is applicable to various data regressions with a large number of zero values in the dependant variable.

## Declaration of Competing Interest

There are no conflicts of interest for the author of this paper.

## Acknowledgement

Not applicable.